\documentclass[11pt]{article}
\usepackage{amsfonts}
\usepackage{amsmath,amssymb}
\usepackage{graphicx}

\baselineskip %
\advance %
\textheight by %
\topskip %
\makeatletter \@addtoreset{equation}{section}

\makeatother
\def\Z{\mathbb Z}

\newcommand{\half}{{\scriptstyle{\frac{1}{2}}}}

\def\be{\begin{equation}}
\def\ee{\end{equation}}

\def\bea{\begin{eqnarray}}

\def\obj(#1)(#2)#3#4#5{%
  \psline[arrows={[-]}, linestyle=dashed, dash=0.0 1,dashadjust=false](#1)(#2)%
  \uput{0.4}[90](#1){#3}%
  \uput{0.4}[-90](#1){#4}\uput{0.4}[-90](#2){#5}%
}
\def\objd(#1)(#2)#3#4#5{%
  \psline[arrows={[-}, linestyle=dashed, dash=0.0 1,dashadjust=false](#1)(#2)%
  \uput{0.4}[90](#1){#3}%
  \uput{0.4}[-90](#1){#4}\uput{0.4}[-90](#2){#5}%
}

\def\eea{\end{eqnarray}}

\def\ben{\begin{displaymath}}
\def\een{\end{displaymath}}
\def\ba{\begin{array}{c}}
\def\bal{\begin{array}{l}}
\def\ea{\end{array}}

\begin{document}

%%%%%%%%%%%%%%%%%%%%%%%%%%%%%%%%%%%%%%%%%%%%%%%%%%%%%%%%%%%%%%%%%%%%%%%%%%%%%%
\title{Super-extended noncommutative Landau problem and conformal symmetry}
%%%%%%%%%%%%%%%%%%%%%%%%%%%%%%%%%%%%%%%%%%%%%%%%%%%%%%%%%%%%%%%%%%%%%%%%%%%%%%

\author{
{\sf Pedro D. Alvarez${}^a$}, {\sf Jos\'e L. Cort\'es${}^b$}, {\sf
Peter A. Horv\'athy${}^c$} {\sf\ and Mikhail S. Plyushchay${}^a$}
\\[4pt]
{\small \it ${}^a$Departamento de F\'{\i}sica, Universidad de
Santiago de Chile, Casilla 307, Santiago 2, Chile}\\
{\small \it ${}^b$Departamento de F\'{\i}sica Te\'orica, Universidad de Zaragoza, Zaragoza 50009, Spain}\\
{\small \it ${}^c$Laboratoire de Math\'ematiques et de
Physique Th\'eorique, Universit\'e de Tours,}\\
{\small \it Parc de Grandmont,
 F-37200 Tours, France}\\
 {\small \it E-mails: pd.alvarez.n@gmail.com, cortes@unizar.es, horvathy@lmpt.univ-tours.fr, mplyushc@lauca.usach.cl}}

\date{}%draft26}

\maketitle

\begin{abstract}
A supersymmetric spin-1/2 particle in the noncommutative plane,
subject to an arbitrary magnetic field, is considered, with
particular attention paid to the homogeneous case.
 The system has
three different phases, depending on the magnetic field. Due to
supersymmetry, the boundary critical phase which separates the sub-
and super-critical cases can be viewed as a reduction  to the
zero-energy eigensubspace. In the sub-critical phase the system is
described by the superextension of exotic Newton-Hooke symmetry,
combined with the conformal $so(2,1)\sim su(1,1)$ symmetry; the
latter is changed into $so(3)\sim su(2)$ in the super-critical
phase. In the critical phase the spin degrees of freedom are frozen
and supersymmetry disappears.
\end{abstract}

%%%%%%%%%%%%%%%%%%%%%%
\section{Introduction}
%%%%%%%%%%%%%%%%%%%%%%

The Landau problem in the noncommutative plane belongs to a family
of non-relativistic systems with so called ``exotic'' symmetries,
meaning that the symmetry algebras have two central charges.  This
is possible only in $d=2+1$ dimensions; for $d\neq 2+1$, the mass is
the unique central charge \cite{Bacry:1968zf} \footnote{In the
relativistic case all central extensions are trivial (central
charges can be absorbed by redefining the generators) \cite{AzHer}.
}. Examples of doubly centrally extended symmetries are provided by
exotic Galilei
\cite{LL,BGO,BGGK,LSZ,DH,Horvathy:2002vt,Horvathy:2004fw}, and
exotic Newton-Hooke (ENH)  \cite{Mariano,Gao,Alvarez:2007fw}
symmetries. In both cases, the second central charge corresponds to
the non-commutativity of the boost generators. The non-commutative
Landau problem (NCLP) carries, in particular, an exotic Newton-Hooke
symmetry \cite{NH-NCL}, which becomes  exotic Galilean symmetry in
the free limit.

The NCLP was investigated in the context of the quantum Hall effect
\cite{DH}, and the physics of anyons \cite{HPany}. In the spinless
case it possesses three different phases \cite{HPany} \footnote{Sub-
and super-critical phases were discussed before in Ref. \cite{BNS}.
The framework used there, however, is only consistent if the
magnetic field is homogeneous, see \cite{Horvathy:2002wc}. Neither
conformal nor supersymmetric extensions are considered in that
paper.}, namely a sub- ($\beta<1$) and a super- ($\beta>1$) critical
ones, separated by a critical phase when $\beta=1$. Here
$\beta=\theta B$ the product
 of the (homogeneous) magnetic field $B=const$ and the
non-commutativity parameter $\theta$ is a dimensionless parameter.
The critical phase is
characterized by a loss of degrees of freedom.
In the generic case of inhomogeneous magnetic
field $B(x)$, all three  phases can simultaneously  be present
in  total phase space.

Non-relativistic conformal symmetry
\cite{NRConf,Duval:1993hs,HMS} is attracting
much current attention, particularly in the context of the AdS/CFT
correspondence
\cite{LeivaPl,Balasubramanian:2008dm%Son:2008ye,Duval:2008jg
}. The usual ($\theta=0$) Landau problem has non-relativistic
conformal symmetry. This raises the following question: {\em What
happens with this symmetry in the noncommutative case?} A priori,
the answer is not obvious. The reason is that conformal symmetry
involves scale invariance, while the noncommutativity parameter,
$\theta$, introduces an independent length scale, $[\theta]=\ell^2$,
in addition to those associated with the mass parameter and the
magnetic field. Scale invariance, nevertheless, could be expected in
the sub-critical phase, whose properties are, in many aspects,
similar to those of the usual Landau problem with $\theta=0$. But it
is not clear what  happens with this symmetry in the super-critical
phase, and for a spin-1/2 particle in particular.

The commutative Landau problem for a non-relativistic electron with
spin $\half $ has an $N=2$ supersymmetry \cite{LandauSUSY}, see also
\cite{CIMT}, and the conformal $so(2,1)$ symmetry is extended into a
superconformal $osp(2|2)$ symmetry, with the angular momentum as
central charge \cite{Duval:1993hs,AnPl}. However, in the NCLP, the
angular momentum behaves in an essentially different way in the sub-
and super- critical phases: it takes values of both signs for
$\beta<1$, but values of only one sign  (depending on the sign of
magnetic field) when $\beta>1$ \cite{HPany}.

The present paper is devoted to the investigation of the
symmetries of the super-extended NCLP for a spin-1/2 particle of
gyromagnetic ratio $g=2$~\footnote{For previous works on
non-commutative supersymmetry the reader is referred to
\cite{Khare}--\cite{Geloun}, where the superextension of exotic
Galilei and Schr\"odinger symmetries, and some aspects of
superextended NCLP were discussed. Questions related to the
super-extension of conformal and exotic Newton-Hooke symmetries
for the NCLP were not considered there.}.

The paper is organized as follows. In Section 2 we start with
 an arbitrary magnetic field and in Section 3 we restrict our
considerations to a homogeneous one. We
discuss the three phases of the system and the differences between them.
Section 4 is devoted to the investigation of the super-symmetries of the system,
related to  exotic Newton-Hooke symmetry and its
extension by adding dilatations and expansions. In Section
5 we switch to an alternative basis of quadratic generators
that allow us to reveal the essential difference between the
sub- and
super- critical phases from the viewpoint of conformal
symmetry. Section 6 summarizes the results.

%%%%%%%%%%%%%%%%%%%%%%%%%%%%%%%%%%%%%%%%%%%%%%%%%%%%%%%%%%%%%%%%%%%%%%%%%%%%%%%%%%%%
%%%%%%%%%%%%%%%%%%%%%%%%%%%%%%%%%%%%%%%%%%%%%%%%%%%%%%%%%%%%%%%%%%%%%%%%%%%%%%%%%%%%%%%%%
\section{$N=2$ supersymmetry: arbitrary magnetic field}
%%%%%%%%%%%%%%%%%%%%%%%%%%%%%%%%%%%%%%%%%%%%%%%%%%%%%%%%%%%%%%%%%%%%%%%%%%%%%%%%%%%%%%%%%%%%%%%

The first order Lagrangian of a spinless ``exotic''  particle
in an arbitrary planar magnetic field $B=B(x)$ \cite{DH,NH-NCL},
\begin{equation}
    L=P_i \dot{x}_i - \frac{1}{2m}
    P_i^2-\frac{\theta}{2}\epsilon_{ij}\dot{P}_iP_j-
    \frac{B}{2}\epsilon_{ij}\dot{x}_ix_j,
\end{equation}
corresponds, at the quantum level, to the commutation relations,
\begin{equation}
    \left[ x_{i},x_{j}\right] =i\frac{\theta }{1-\beta }\epsilon _{ij},
    \qquad \left[ x_{i},P_{j}\right]
    =i\frac{1}{1-\beta }\delta _{ij},\qquad
    \left[ P_{i},P_{j}\right] =i\frac{B}{1-\beta}\epsilon
    _{ij},
    \label{conmutadores1}
\end{equation}
where $\epsilon _{ij}$ is the antisymmetric tensor with $\epsilon
_{12}=1$ \footnote{We have chosen units $\hbar=1=c$ and put the
electric charge equal to one.}. The parameter $\theta $ is related
to the noncommutativity of the coordinates and has the dimension of
a squared length $\ell^2$. The dimension of the magnetic field,
$B(x)$, is $\ell^{-2}$.  $\beta=\beta(x)=\theta B(x)$ is then
dimensionless. Spin degrees of freedom are introduced by
supplementing (\ref{conmutadores1}) with
\begin{equation}
    \left\{ S_{i},S_{j}\right\} =\frac{1}{2}\delta _{ij},
    \qquad \left[ S_{i},x_{j}\right] =0,\qquad
    \left[ S_{i},P_{j}\right] =0. \label{conmutadores2}
\end{equation}
The coordinates $x_{i}$ and momenta $P_{i}$ are bosonic, while the
spin-1/2 operators $S_{i}$ are fermionic variables. The relations
(\ref{conmutadores1})-(\ref{conmutadores2}) specify a consistent
quantum structure  (Jacobi identities hold for an arbitrary
inhomogeneous magnetic field \cite{Horvathy:2002wc}).

 In the critical case $\beta=1$
the bosonic differential two-form associated with
(\ref{conmutadores1}) becomes degenerate, and, in the spinless case,
it requires  special consideration \cite{DH,HPany}. In the
superextended NCLP we consider below, the critical phase  $\beta=1$
will be obtained by reducing the system to the (infinitely
degenerate) zero energy subspace.

The fermionic operators
\begin{equation}
    Q_{1} =\sqrt{\frac{2}{m}}\,P_{i}S_{i},
    \qquad Q_{2} =\sqrt{\frac{2}{m}}\,\epsilon _{ij}P_{i}S_{j}
    \label{Q's}
\end{equation}
generate  an $sl(1|1)$ superalgebra \cite{KhareSukhatme},
\begin{equation}\label{Hgen}
    \left\{ Q_{a},Q_{b}\right\}=2H\delta_{ab} ,
    \qquad
    \left[ Q_{a},H\right]=0,
\end{equation}
where
\begin{equation}
    H=\frac{1}{2m}P_{i}^{2}-\omega S_{3},
    \label{H}
%\end{equation}
%\begin{equation}\label{omBm}
\qquad
    \omega=\frac{B}{m^*},\qquad
     m^*=m(1-\beta),
\end{equation}
and $S_3$ is defined by
\begin{equation}\label{S3}
    S_{3}=-i\epsilon _{ij}S_{i}S_{j}.
\end{equation}
Choosing $H$ as the Hamiltonian, we get a system that generalizes
the usual $N=2$ supersymmetry of a spin-$1/2$ particle with
gyromagnetic ratio $g=2$ in arbitrary magnetic field
\cite{KhareSukhatme} to the non-commutative case. The operator
$S_3$, like the supercharges $Q_a$, is an integral of the motion,
which acts  as a  $\Z_2$-grading operator $\Gamma$ for the $N=2$
supersymmetry,  $\Gamma=2S_3$, $\Gamma^2=1$. One can choose, in
particular, a representation where $S_3$ is proportional to the
diagonal Pauli matrix, $S_3=\frac{1}{2}\sigma_3$.

%%%%%%%%%%%%%%%%%%%%%%%%%%%%%%%%%%%%%%%%%%%%
\section{Three phases of the superextended NCLP}
%%%%%%%%%%%%%%%%%%%%%%%%%%%%%%%%%%%%%%%%%%%%

{}From now on we consider a homogeneous field $B=const\neq0$. It is
convenient to define a linear combination of the bosonic operators
$x_i$ and $P_i$,
\begin{equation}\label{calP}
    {\cal P}_i=P_i-B\epsilon_{ij}x_j.
\end{equation}
For nonzero magnetic field, the set formed by the ${\cal P}_i$ and
$P_i$ is an alternative to the initial set of bosonic variables. The
advantage is that the ${\cal P}_i$ commute with the $P_j$. {}From
the form of the Hamiltonian (\ref{H}) we infer that ${\cal P}_i$ is
an integral of the motion. Since $[x_i,{\cal P}_j]=i\delta_{ij}$ and
\begin{equation}
\label{calPP}
    \left[ \mathcal{P}_{i},\mathcal{P}_{j}\right] =
    -iB\epsilon_{ij},
\end{equation}
the ${\cal P}_i$, $i=1,2$, are identified as the non-commuting
generators of space translations. Another bosonic operator,
\begin{eqnarray}
    J=\frac{1}{2B}\left({\cal P}_i^2-(1-\beta)P_i^2\right)+S_3
    =
    \frac{B}{2}\left(x_i+\frac{1}{B}
    \epsilon_{ij}P_j\right)^2-\frac{1-\beta}{2B}P_i^2+S_3,
    \label{J(x,P)}
\end{eqnarray}
is identified as the angular momentum, since it generates
the rotations,
$[J,R_i]=i\epsilon_{ij}R_j$ for $R_i=x_i,\, P_i,\, S_i$.

Putting $\varepsilon_z=\mathrm{sgn}(z)$, we define bosonic and
fermionic creation and annihilation operators
\begin{eqnarray}
    &a^{+}=(a^-)^\dagger=\sqrt{\frac{|1-\beta|}{2\left\vert B \right\vert
    }} \left( P_{1}- i \varepsilon_{B}\varepsilon_{\left( 1-\beta
    \right)} P_{2}\right) ,\qquad
    b^{+}=(b^-)^\dagger=\frac{1}{\sqrt{2\left\vert B\right\vert }}\left(
    \mathcal{P}_{1}+i \varepsilon_B \mathcal{P}_{2}\right),&\label{ab+-}\\
    &f^{+}=(f^-)^\dagger=S_{1}- i \varepsilon_{B}\varepsilon_{\left( 1-\beta \right)}
    S_{2}.&
    \label{f+-}
\end{eqnarray}
Their nonzero (anti)-commutation relations are
\begin{equation}
    \left[ a^{-},a^{+}\right] =1,\qquad \left[ b^{-},b^{+}\right] =1,
    \qquad \left\{ f^{-},f^{+}\right\}
    =1.
    \label{crea-anhi}
\end{equation}
The  definition of the bosonic creation and annihilation
operators depends, in view of (\ref{conmutadores1}) and
(\ref{calPP}), on the signs of the magnetic field and of the quantity
$1-\beta$ that defines the phase of the system.
The dependence is included into the definition of the fermionic operators. This
allows us to present the Hamiltonian in both the sub- ($\beta<1$) and
super- ($\beta>1$) critical phases in a universal form,
\begin{equation}\label{HNN}
    H=\left\vert \omega \right\vert
    \left( N_{a}+N_{f}\right).
\end{equation}
Here we introduced the bosonic,
$N_a=a^+a^-$ and  $N_b=b^+b^-$, and  fermionic,
 $N_f=f^+f^-$,  number operators with eigenvalues  $n_a,
n_b=0,1,\ldots$, and $n_f=0,1$, respectively.
The angular momentum reads,
\begin{equation}\label{JNN}
    J=\varepsilon_B \left( N_{b}+\frac{1}{2}\right)
    -\varepsilon_{ B}\varepsilon_{\left( 1-\beta \right)}\left(
    N_{a}+N_{f}\right).
\end{equation}

According to (\ref{HNN}), in both non-critical phases the system
has a typical $N=2$ supersymmetric spectrum with zero ground state
energy corresponding to $n_a=n_f=0$, and supersymmetric energy
doublets
%given by the pairs of the
with quantum numbers $n_a>1$, $n_f=0$,  and $n_a-1$, $n_f=1$,
respectively. Each energy level has an additional infinite
degeneracy  ($n_b=0,1,\ldots$), associated with the translation
invariance generated by the ${\cal P}_i$.

 On the other hand, Eq. (\ref{JNN}) reveals the essential
difference between two non-critical phases. In the subcritical
phase, the angular momentum takes half-integer values of any sign,
while in super-critical phase it only takes half-integer values of
one sign
%, which coincides with
(the sign of the magnetic field).

It is worth noting that the difference between the two
non-critical phases reveals itself also in another aspect.
Proceeding from the quantum structure (\ref{conmutadores1}), one
can construct  \emph{vector} variables $q_i$ and $p_i$ with
canonical commutation relations $[q_i,q_j]=[p_i,p_j]=0$,
$[q_i,p_j]=i\delta_{ij}$. Up to a unitary transformation, they can
be presented in a simple form in terms of the mutually commuting
operators ${\cal P}_i$ and $P_i$,
\begin{equation}
    q_i=\frac{1}{B}\epsilon_{ij}\left({\cal P}_j-\sqrt{1-\beta}\,P_j\right),
    \qquad
    p_i=\frac{1}{2}\left({\cal P}_i+\sqrt{1-\beta}\, P_i\right).
    \label{qp}
\end{equation}
In the limit $B\rightarrow 0$, $q_i$ becomes the canonical
coordinate for a free particle on the non-commutative plane,
$q_i=x_i+\frac{\theta}{2}\epsilon_{ij}P_j$, and $p_i=P_i$, see Refs.
\cite{DH,Horvathy:2002vt,Horvathy:2004fw}. Eq. (\ref{qp}) provides
us with canonical coordinates and momenta both in the sub- and
supercritical phases. However, in the super-critical case, unlike
in the sub-critical phase, the operators (\ref{qp}) are
\emph{non-hermitian}.

Consistently with Eq. (\ref{H}), in the limit $\beta\rightarrow 1$
the frequency tends to infinity, $|\omega|\rightarrow \infty$. The
critical phase $\beta=1$ can be obtained by reduction of the system
to the lowest energy level $E=0$, where $n_a=n_f=0$. In this phase
the system is described by the oscillator variables $b^\pm$
(non-commuting translation generators ${\cal P}_i$), and $H=0$.
Taking into account Eqs. (\ref{S3}) and (\ref{f+-}), we find that
$S_3=\frac{1}{2}\varepsilon_B\varepsilon_{(1-\beta)}$, i.e. , the
spin projection is fixed. Curiously, its value depends on the phase
from which the reduction is realized.
 In the critical phase the spin
degrees of freedom, like those associated with the bosonic
oscillator variables $a^\pm$, are frozen, and supersymmetry
disappears.

The Virasoro algebra can be realized in terms of the remaining
bosonic integrals $b^\pm$ \cite{Kumar}.
Restricting ourselves
 to integrals  of
degree not higher than $2$ in the operators $b^\pm$,
provides us with the symmetry algebra of the planar Euclidean group, spanned
 by the angular momentum,
$J=\varepsilon_{B}(N_b+\frac{1}{2})$, and by the
non-commuting translation
generators ${\cal P}_i$, see Eqs. (\ref{calPP}) and
(\ref{crea-anhi}).

In what follows we suppose $\beta\neq 1$.

%%%%%%%%%%%%%%%%%%%%%%
\section{Symmetries}
%%%%%%%%%%%%%%%%%%%%%%

Here we identify the other symmetries of the system in addition to
those described in the previous section. For this purpose, we
consider the Hamiltonian equations of motion,
\begin{equation}
    \label{xpsevol}
    \dot{x}_{i}=\frac{1}{m^*}P_{i},
    \qquad
    \dot{P}_{i}=\omega \epsilon _{ij}P_{j},
    \qquad
    \dot{S}_{i}=\omega \epsilon _{ij}S_{j}.
\end{equation}
In the sub- and super- critical phases the evolution is of the same
form, but (assuming a given sign for the field $B$) the sign of the
effective mass, $m^*$, (and of the frequency, $\omega$,) is opposite
for $\beta<1$ and $\beta>1$. Remarkably, the same effect can be
produced by a time reflection, $t\rightarrow -t$. The integration of
the equations of motion gives,
\begin{equation}
     x_{i}(t) = \frac{1}{B}\left(\epsilon_{ij}{\cal P}_j
    -\Delta _{jk}^{-1}(t) P_{k}(0)\right) ,\quad
    P_{i}(t) = \Delta _{ij}^{-1}(t) P_{j}(0) ,\quad
    S_{i}(t) = \Delta _{ij}^{-1}(t) S_{j}(0) .
\end{equation}
Here $\Delta_{ij}(t)=\cos{\omega t} \, \delta_{ij}-\sin{\omega t}
\, \epsilon_{ij}$ is a rotation matrix,
$\Delta_{ij}^{-1}(t)=\Delta_{ji}(t)=\Delta_{ij}(-t)$ is its
inverse, and $\frac{1}{B}\epsilon_{ij}{\cal
P}_j=\frac{1}{B}\epsilon_{ij}{\cal
P}_j(0)=x_i(0)+\frac{1}{B}\epsilon_{ij}P_{j}(0)$.

 Now, we identify
the boost generators  as the integrals which, when
acting on $x_i(0)$ and $\dot{x}_i(0)$, produce the necessary form of
the infinitesimal transformations, $[x_{i}(0) ,\mathcal{K}_{j}]
=0,$ $[ \dot{x}_{i}( 0) ,\mathcal{K} _{j}] =-i\delta _{ij}$. This
gives $\mathcal{K}_{i}=m^{*}\left( x_{i}\left( 0\right) +\theta
\epsilon _{ij}P_{j}\left( 0\right) \right)$. Using the  solution
of the equations of motion, the generators can be rewritten in
terms of the variables  $x_i$ and $P_i$,
\begin{equation}
    \mathcal{K}_{i}=m^{*}\left(x_{i}+\frac{1}{B}
    \epsilon_{ij}\rho_{jk}P_{k}\right),\qquad
    \rho_{jk}\equiv \left( \delta_{jk}-\left(1-\beta\right)
    \Delta _{jk}\left( t\right) \right).
    \label{K_i}
\end{equation}
The boost generators (\ref{K_i}) are dynamical integrals of motion
in the sense that they explicitly depend on time,
$\frac{d}{dt}{\cal K}_j=\frac{\partial}{\partial t}{\cal
K}_j+\frac{1}{i}[{\cal K}_j,H]=0$.

 For $\theta=0$,   (\ref{K_i})
reproduces correctly the boost generators of the usual Landau
problem \cite{OP}. In the free case $B=0$, (\ref{K_i}) reduces to
the  boost generators of the free particle in the non-commutative
plane.

The commutators between $\mathcal{P}_{i}$ and $\mathcal{K}_{i}$
are given by
\begin{equation}
    \left[ \mathcal{K}_{i},\mathcal{K}_{j}\right] =-i\theta m^{\ast
    2}\epsilon _{ij},\qquad \left[
    \mathcal{K}_{i},\mathcal{P}_{j}\right] =im^{\ast }\delta
    _{ij},\qquad \left[ \mathcal{P}_{i},\mathcal{P}_{j}\right]
    =-im^*\omega\epsilon _{ij},  \label{K_P}
\end{equation}
and their commutation relations with $H$ are
\begin{equation}
    \left[ \mathcal{P}_{i},H\right] =0,\qquad \left[ \mathcal{K}_{i},H\right] =i\left(
    \mathcal{P}_{i}+\omega \epsilon _{ij}\mathcal{K}_{j}\right). \label{P_K_H}
\end{equation}
The  bosonic integrals $H$, $J$, $\mathcal{P}_{i}$ and
$\mathcal{K}_{i}$ generate the  exotic Newton-Hooke symmetry
algebra, in which $\omega$ is a parameter, while
$C=m^{\ast }$ and $\tilde{C}=\theta m^{\ast 2}$ are central charges
\cite{NH-NCL}. The commutators with the supercharges $Q_a$ show that
they are translation-, but not boost-invariant,
\begin{equation}
    \left[ \mathcal{P}_{i},Q_{a}\right] =
    0,\qquad \left[ \mathcal{K} _{i},Q_{a}\right]=i\left(
    \delta_{a1}\Sigma
    _{i}+\delta_{a2}\epsilon _{ij}\Sigma _{j}\right).
    \label{P,K-Q}
\end{equation}
Here we have identified a new, fermionic vector generator $\Sigma
_{i}=\left( 1-\beta \right) \sqrt{2m}S_{i}\left( 0\right)$. This is
again a dynamical integral,
\begin{equation}
    \Sigma _{i}=\left( 1-\beta \right)
    \sqrt{2m}\Delta _{ij}\left( t\right) S_{j}.
\end{equation}
{}The anticommutation relations
\begin{equation}
    \left\{ \Sigma _{i},\Sigma _{j}\right\} ={\cal C}\delta _{ij},\qquad
    {\cal C}=C-\omega \tilde{C}=m(1-\beta)^2>0,
    \label{sigma-sigma}
\end{equation}
imply that $\Sigma_i$ is the square root of a suitable
positive definite linear combination of the central charges.

The commutation relations of $\Sigma _{i}$
 with $\mathcal{K}_{i}$, $\mathcal{P}_{i}$, $J$ and $H$ are
\begin{equation}
    \left[\Sigma _{i},\mathcal{K}_{j}\right] =0,
    \qquad \left[ \Sigma_{i}, \mathcal{P}_{j}\right] =0,\qquad
    [J,\Sigma_i]=i\epsilon_{ij}\Sigma_j,\qquad
    \left[ \Sigma _{i},H\right] =i\omega
    \epsilon _{ij}\Sigma _{j}.
    \label{Sigma-H}
\end{equation}
As it follows from (\ref{P,K-Q}), the $\Sigma_i$ inherit the explicit
time  dependence of the $\mathcal{K}_i$. The anticommutators
with the supercharges $Q_a$ are
\begin{equation}
    \left\{ \Sigma _{i},Q_{a}\right\}
    =\left(\delta_{a1}\delta_{ij}-\delta_{a2}\epsilon_{ij}\right)
    \left(
    \mathcal{P}_{j}+\omega \epsilon _{jk}\mathcal{K}_{k}\right) .
\end{equation}
The integrals $H$, $J$, $\mathcal{P}_i$, $\mathcal{K}_i$, $Q_a$ and
$\Sigma_i$ generate a closed Lie-type superalgebra centrally
extended by $C$ and $\tilde{C}$, in which the frequency, $\omega$, plays
the role of a parameter.

Let us now  inquire about conformal symmetry. To identify its
generators, we first consider the direct analogs of the dilatation
and special conformal symmetry (expansion) generators of a free
particle \cite{LeivaPl},
\begin{equation}
    D=\frac{1}{4m^{\ast }}\left( \mathcal{K}_{i}\mathcal{P
    }_{i}+\mathcal{P}_{i}\mathcal{K}_{i}\right) ,\qquad
    K=\frac{1}{2m^{\ast } }\mathcal{K}_{i}^{2}.  \label{DK}
\end{equation}
The commutation relations,
\begin{eqnarray}
    \left[K,H\right]= 2iD,\qquad
    \left[ D,H\right]=\frac{i}{2}
    \Big(\left( 2-\beta\right)H+\omega
    \left(J+\left( 1-\beta\right)S_{3}-\omega K\right) \Big),
    \label{D-H}
\\[6pt]
    \left[K,D\right]  = \frac{i}{2}
    \Big( \left( 2-\beta \right) K-m^*\theta
    \left( J-\left( 1-\beta \right) S_{3}+m^*\theta H\right) \Big),  \label{K-D}
\end{eqnarray}
generalize the ``exotic" relations found before for a free spinless
``Moyal'' field \cite{HMS}.

In contrast with the usual spinless and free case $\theta=B=0$, the
commutators do not close to an $so(2,1)$ algebra. But, since $J$ and
$S_3$ commute with $H$, $D$ and $K$, one could conclude that we have
a kind of central extension of $so(2,1)$. As we shall see below,
this is only true in the sub-critical phase, while in the
super-critical phase the noncompact $so(2,1)$ algebra is changed
into the compact $so(3)$. Since $\omega$ plays a role of a parameter
in Newton-Hooke symmetry, in the commutative case $\theta=0$ the
relations (\ref{D-H}), (\ref{K-D}) correspond to an $so(2,1)$
algebra, centrally extended by $J+S_3$. In non-commutative case
$\theta\neq 0$, however, we do not have a Lie-algebraic structure,
due to the dependence of the coefficients in (\ref{D-H}) and
(\ref{K-D}) on the central charges of exotic Newton-Hooke symmetry.
In the next section we will consider an alternative choice of the
generators
 that linearizes a superalgebraic structure.

To identify the complete
superalgebraic structure, we will also need the commutators of
$D$ and $K$ with the other generators of the super-extended exotic
Newton-Hooke symmetry. The commutators with $\mathcal{P}_{i},\
\mathcal{K}_{i}$ and $\Sigma _{i}$ are
\begin{equation}\label{KPKSig}
    \left[K, \mathcal{P}_{i}\right]  =i\mathcal{K}_{i},
    \qquad
    \left[K, \mathcal{K}_{i}\right]  =im^*\theta\epsilon _{ij}\mathcal{K}_{j},\qquad
    \left[K, \Sigma _{i}\right]  =0,
\end{equation}
\begin{equation}\label{DPK1}
    \left[D, \mathcal{P}_{i}\right]  =\frac{i}{2 }
    \left( \mathcal{P}_{i}+\omega \epsilon _{ij}\mathcal{K}_{j}\right)
    ,\qquad
    \left[D, \mathcal{K}_{i}\right]
    =\frac{i}{2}m^*\theta\epsilon_{ij}
    \left(\mathcal{P}_{j}+\omega \epsilon _{jk}\mathcal{K}_{k}\right)
    ,\qquad
    \left[D, \Sigma _{i}\right]  =0,
\end{equation}
where, again, the nonlinearity is manifest. The commutators with the
 supercharges $Q_a$,
\begin{equation}\label{QKD}
    \left[K, Q_a\right] =iZ_a,\qquad
    \left[D, Q_a\right] =\frac{i}{2}
    \Big( (1-\beta)Q_a+\omega \epsilon
    _{ab}Z_b\Big),
\end{equation}
generate a new set of the scalar  supercharges $Z_a$,
\begin{equation}\label{Zdef}
    Z_{1} =\frac{1}{m^{\ast } }
    \mathcal{K}_{i}\Sigma_{i}, \qquad
    Z_{2} =\frac{1}{m^{\ast } }
    \epsilon _{ij}\mathcal{K} _{i}\Sigma _{j},
\end{equation}
with anticommutators
\begin{equation}
    \left\{ Z_{a },Z_{b }\right\} =2 (1-\beta)
    \delta _{ab}\left( K+m^*\theta
    S_{3}\right).
\end{equation}
The (anti)-commutation relations of $Z_a$ with  other symmetry
generators,
\begin{equation}
    \left[ Z_{a },H\right] =iQ_{a},\qquad
    \left[ Z_{a},K\right] =im^*\theta
    \epsilon _{ab }Z_{b },\qquad
    \left[ Z_{a },D\right] =\frac{i}{2}(1-\beta)\left( Z_{a}+m^*\theta
    \epsilon _{ab }Q_{b}\right),
\end{equation}
\begin{equation}
    \left[Z_a, \mathcal{P}_{i}\right]  = i\left(
    \delta_{a1}\delta_{ij}+\delta_{a2}\epsilon
    _{ij}\right)\Sigma _{j}, \qquad
    \left[Z_a, \mathcal{K}_{i}\right]  = im^*\theta \left(\delta_{a1}
    \epsilon _{ij}-\delta_{a2}\delta_{ij}\right)\Sigma _{j},
    \label{ZZ1}
\end{equation}
\begin{equation}
    \left\{Z_a, \Sigma _{i}\right\}  = (1-\beta)\left(\delta_{a1}
    \delta_{ij}-\delta_{a2}\epsilon _{ij}\right)
    \mathcal{K}_{j},
    \qquad
    \left\{Z_a, Q_{b}\right\} =2D\ \delta _{ab}-
    \left( J+\left( 1-\beta \right) S_{3}-\omega
    K +m^*\theta H\right)\epsilon _{ab},
    \label{Q,Z}
\end{equation}
show that  a closed super-algebraic structure is obtained, and no
new independent integrals are generated.  In the commutative case
$\theta=0$ this super-algebraic structure reduces to the
Schr\"odinger superalgebra studied in \cite{Duval:1993hs}.

%%%%%%%%%%%%%%%%%%%%%%%%%%%
\section{Alternative basis}
%%%%%%%%%%%%%%%%%%%%%%%%%%%

In this section we show that, changing the basis of the
 conformal symmetry generators, the nonlinearity due to
the presence of central charges in the coefficients in the
(anti)commutation relations can be removed, and we get a certain
Lie-superalgebraic extension of the conformal symmetry.
 The linearization procedure can be extended to include also the
generators of translations and boosts, and the vector supercharge
$\Sigma_i$.
%In detail,
We consider
%\begin{eqnarray}
\begin{equation}
    \mathcal{J}^0=\frac{1}{\omega}H+\frac{1}{2}(J+S_3),
    \quad
    \mathcal{J}^1=\frac{\varepsilon_{1-\beta}}{2\sqrt{|1-\beta|}}
    \left(\frac{2-\beta}{\omega}H+J+(1-\beta)S_3-\omega
    K\right),\quad
    \mathcal{J}^2=\frac{1}{\sqrt{|1-\beta|}}\,D,
    \label{J2}
\end{equation}
instead of $H$, $D$, $K$. Note that ${\cal J}^1$ and ${\cal J}^2$
depend nontrivially on the noncommutative parameter $\theta$ via
$\beta$. All three  integrals (\ref{J2}) depend only on the bosonic
variables $a^\pm$, $b^\pm$ but not on the fermionic operators
$f^\pm$,
\begin{eqnarray}
    &\mathcal{J}^0=\frac{1}{2}\varepsilon_B(N_b+\varepsilon_{1-\beta}N_a)+
    \frac{1}{4}\varepsilon_B(1+\varepsilon_{(1-\beta)}),&
    \\[6pt]
    &\mathcal{J}^1_{sub}=\frac{1}{2}\varepsilon_B\left(a^+(0)b^++a^-(0)b^-\right),\qquad
    \mathcal{J}^2_{sub}=\frac{-i}{2}\left(a^+(0)b^+-a^-(0)b^-\right),&
    \\[6pt]
    &\mathcal{J}^1_{sup} =
    \frac{1}{2}\varepsilon_B\left(a^+(0)b^-+a^-(0)b^+\right),\qquad
    \mathcal{J}^2_{sup} =
    \frac{i}{2}\left(a^+(0)b^--a^-(0)b^+\right),&
\end{eqnarray}
where the subscripts \emph{sub} and \emph{sup} refer to the sub- and
super- critical phases, and $a^\pm(0)=a^\pm e^{\mp i|\omega|t}$. The
commutation relations of ${\cal J}^\mu$, $\mu=0,1,2$, are given by
\begin{equation}\label{so}
    \left[ \mathcal{J}^{1},\mathcal{J}^{2}\right]
    =-i\varepsilon _{1-\beta } \mathcal{J}^{0},\qquad
    \left[ \mathcal{J}^{2},\mathcal{J}^{0}\right]
    =i \mathcal{J}^{1},\qquad \left[
    \mathcal{J}^{0},\mathcal{J}^{1}\right] =i \mathcal{J}^{2}.
\end{equation}
In the sub-critical case this is the $so(2,1)\sim su(1,1)$
algebra, but in the supercritical case the operators (\ref{J2})
generate the $so(3)\sim su(2)$ algebra.

Together with the new basis of bosonic generators, we define the
following linear combinations of the scalar supercharges $Q_a$ and
$Z_a$,
\begin{equation}\label{comsusy1}
    {\cal Q}_a^+=\frac{1}{2|\omega|}
    \left((1+\sqrt{|1-\beta|})Q_a+\frac{\omega}{\sqrt{|1-\beta|}}\epsilon_{ab}Z_b\right),\, \,
    {\cal Q}_a^-=\frac{1}{2|\omega|}\epsilon_{ab}\left((1-\sqrt{|1-\beta|})Q_b
    -\frac{\omega}{\sqrt{|1-\beta|}}\epsilon_{bc}Z_c\right).
\end{equation}
Then, in the sub-critical phase, we get the (anti)-commutation relations
\begin{equation}\label{VV WW sub}
    \left\{ {\cal Q}_a^+,{\cal Q}_b^+\right\} =2\varepsilon_{B}\delta_{ab}
    \left(\mathcal{J} ^{0}+\mathcal{J}^{1}\right), \
    \qquad \left\{ {\cal Q}_a^-,{\cal Q}_b^-\right\}
    =2\varepsilon_{B}\delta_{ab}\left(\mathcal{J}^{0}-
    \mathcal{J}^{1}\right),
\end{equation}
\begin{equation}\label{VW sub}
    \left\{ {\cal Q}_a^+,{\cal Q}_b^-\right\} =
    2\varepsilon_{B}\left(\delta _{ab} \mathcal{J} ^{2}
    +\epsilon_{ab}\frac{1}{2}\left( J+S_{3}\right)\right),
\end{equation}
\begin{equation}\label{J,V}
    \left[\mathcal{J}^{1},{\cal Q}_a^\pm\right]
    =-\frac{i}{2}{\cal Q}_a^\mp,\qquad
    \left[\mathcal{J}^{2},{\cal Q}_a^\pm\right]=\pm\frac{i}{2}{\cal
    Q}_a^\pm.
\end{equation}
In the super-critical case we instead  find
\begin{equation}\label{VV WW sup}
    \left\{ {\cal Q}_a^+,{\cal Q}_b^+\right\} =
    2\varepsilon _{B}\delta _{ab}
    \left( \frac{1}{2}\left(J+S_{3}\right)
    -\mathcal{J}^{1}\right),
    \qquad \left\{ {\cal Q}_a^-,{\cal Q}_b^-\right\} =
    2\varepsilon _{B}\delta _{ab} \left(
     \frac{1}{2}\left(J+S_{3}\right)
    +\mathcal{J}^{1}\right),
\end{equation}
\begin{equation}
\label{VWsup}
    \left\{ {\cal Q}_a^+,{\cal Q}_b^-\right\} =
    2\varepsilon_{B}\left(\epsilon_{ab}\mathcal{J}^{0}\ -
    \delta _{ab}\mathcal{J}^{2}\right),
\end{equation}
\begin{equation}
    \left[
    \mathcal{J}^{1},{\cal Q}_a^\pm\right]
    =\mp\frac{i}{2}\epsilon _{ab}{\cal Q}_b^\pm,
    \qquad \left[ \mathcal{J}^{2},{\cal Q}_a^\pm\right] =
    \frac{i}{2}\epsilon_{ab}{\cal Q}_b^\mp.
\end{equation}
In both phases, we also have,
\begin{equation}
      \left[ \mathcal{J}^{0},{\cal Q}_a^\pm
    \right] =\pm\frac{i}{2}{\cal Q}_a^\mp,\qquad
    \left[ S_{3},{\cal Q}_a^\pm\right] =
    i\epsilon _{ab}{\cal Q}_b^\pm.
    \label{SVW}
\end{equation}

Let us emphasise that all these relations are linear, as advertised.

Note that the commutators of the supercharges with ${\cal J}^0$
are the same for the sub- and super- critical phases, but those
with ${\cal J}^1$ and ${\cal J}^2$  are different. The relations
(\ref{VV WW sub}), (\ref{VW sub}) and (\ref{VV WW sup}),
(\ref{VWsup}) are transformed mutually under the change ${\cal
J}^0 \leftrightarrow \frac{1}{2}(J+S_3)$, ${\cal J}^1
\leftrightarrow -{\cal J}^1$ and ${\cal J}^2 \leftrightarrow
-{\cal J}^2$.

The relations (\ref{so}) and (\ref{VV WW sub})--(\ref{SVW}) show
that the system is described by the centrally extended
supersconformal $osp(2|2)$ symmetry in the sub-critical phase, and
by the analogous Lie-superalgebraic extension of the compact
$so(3)$ symmetry  in the super-critical phase. In both cases the
angular momentum $J$ plays the role of central charge in these
superalgebras, while ${\cal R}=2S_3$ is the generator of
$R$-symmetry.

Let us take
\begin{equation}\label{defES}
    {\cal E}^+_i=\sqrt{\frac{|1-\beta|}{2|\omega|}}\, {\cal P}_i,\qquad
    {\cal E}^-_i=\frac{1}{\sqrt{2|\omega|}}
    \left({\cal P}_i+\omega\epsilon_{ij}{\cal K}_j\right)
\end{equation}
instead of the translation and boost generators. The nontrivial
(anti)-commutators of the integrals ${\cal E}^\pm_i$ and
$\Sigma_i$ between themselves are given by (\ref{sigma-sigma}) and
\begin{equation}\label{EES}
    [{\cal E}^+_i,{\cal E}^+_j]=-\frac{i}{2}\varepsilon_{B}{\cal C} \epsilon_{ij},
    \qquad [{\cal E}^-_i,{\cal E}^-_j]=\frac{i}{2}
    \varepsilon_{B}\varepsilon_{(1-\beta)}{\cal C} \epsilon_{ij}.
\end{equation}
Their (anti)commutation relations with the rest of generators,
${\cal J}^\mu$, $J$, $S_3$ and ${\cal Q}^\pm_\alpha$, are
\begin{eqnarray}\label{J0ES}
    &[{\cal J}^0,{\cal E}^\pm_i]=\pm\frac{i}{2}\epsilon_{ij}{\cal
    E}^\pm_j,\qquad [J,{\cal E}^\pm_i]=i\epsilon_{ij}{\cal E}^\pm_j,
    \qquad
     [S_3,{\cal E}^\pm_i]=0,&\\
    &[{\cal E}^+_i,{\cal J}^\mu]=-\frac{i}{2}\varepsilon_{(1-\beta)}
    \left(\delta^{\mu 1}\epsilon_{ij}+\delta^{\mu 2}\delta_{ij}\right){\cal
    E}^-_j,
    \quad
    [{\cal E}^-_i,{\cal J}^\mu]=\frac{i}{2}
    \left(\delta^{\mu 1}\epsilon_{ij}-\delta^{\mu 2}\delta_{ij}\right){\cal
    E}^+_j,\,\, \mu=1,2,&
     \label{EJ12}\\
     &[S_3,\Sigma_i]=i\epsilon_{ij}\Sigma_j,
     \qquad
    [J,\Sigma_i]=i\epsilon_{ij}\Sigma_j,\qquad
    [{\cal J}^\mu,\Sigma_i]=0,&\label{SJmu}
\end{eqnarray}
\begin{eqnarray}
    &[{\cal E}^+_i,{\cal Q}^\pm_a]=\frac{i}{2}\varepsilon_{B}\left(
    -\delta_{a1}\epsilon_{ij}\pm \delta_{a2}\delta_{ij}\right)\Sigma_j,\quad
    [{\cal E}^-_i,{\cal Q}^\pm_a]=
    \frac{i}{2}\varepsilon_{B}\varepsilon_{(1-\beta)}\left(
    \pm\delta_{a1}\epsilon_{ij}\mp\delta_{a2}\delta_{ij}\right)\Sigma_j,&\label{comsusy2}\\
    &\qquad \{\Sigma_i,{\cal Q}^+_a\}=\left(\delta_{a1}\delta_{ij}-\delta_{a2}\epsilon_{ij}\right)({\cal
    E}^+_j+{\cal E}^-_j),\quad
    \{\Sigma_i,{\cal Q}^-_a\}=\left(\delta_{a1}\epsilon_{ij}+\delta_{a2}\delta_{ij}\right)({\cal
    E}^+_j-{\cal E}^-_j).\label{comsusy4}&
\end{eqnarray}
The indices of the bosonic generators in (\ref{defES}) correspond to
the signs in their commutators with ${\cal J}^0$ in (\ref{J0ES}).

According to (\ref{sigma-sigma}) and (\ref{EES}), even, ${\cal
E}^\pm_i$,  and  odd, $\Sigma_i$, the integrals generate the
superextended two-dimensional Heisenberg algebra with ${\cal C}$
as central charge, which, in the commutative case
$\theta=0$, becomes the mass, $m$.  This is the unique central
charge of the resulting complete symmetry Lie superalgebra given
by Eqs. (\ref{sigma-sigma}), (\ref{so}), (\ref{VV WW
sub})--(\ref{SVW}), (\ref{EES})--(\ref{comsusy4}).  For
$\theta=0$ it provides us with an alternative form of the Schr\"odinger
superalgebra \cite{Duval:1993hs}.

Let us emphasize that the linearization of the unified
super-extended exotic Newton-Hooke and conformal symmetries is
achieved by inclusion  of the dependence on $\theta$ in the base
changing coefficients. The noncommutativity parameter itself is a
function of the exotic Newton-Hooke symmetry central charges,
$\theta=\tilde{C}/C^2$.

Note that while the scalar supercharges ${\cal Q}^\pm_a$ generate
via anticommutation relations the even scalar integrals ${\cal
J}^\mu$ and $J+S_3$, their anticommutation relations with
$\Sigma_i$ reproduce the even vector generators ${\cal E}^\pm_i$.
The generator ${\cal R}=2S_3$ of the $R$-symmetry is related to
the vector supercharge $\Sigma_i$ via one of the Casimir operators
of the superalgebra, $i\epsilon_{ij}\Sigma_i\Sigma_j+{\cal R}{\cal
C}$, that takes here zero value.

%%%%%%%%%%%%%%%%%%%%%%%%%%%%%%%%
\section{Conclusion and Outlook}

Let us summarize our results.

We  observed that the energy levels in all  three phases are
infinitely degenerate, due to magnetic translation invariance. In
the sub- and super- critical phases nonzero energy levels reveal
also an additional double degeneration, associated with $N=2$
supersymmetry. Due to supersymmetry, the critical boundary phase can
be obtained by a simple reduction of the system to the zero energy
eigenspace. In this phase one bosonic and the spin degrees of
freedom are frozen, and  SUSY disappears. The symmetry associated
with the integrals of degree not higher than two in the residual
bosonic oscillator variables corresponds to the Euclidean group of
motions in two dimensions,  generated by non-commuting magnetic
translations and by rotations. The angular momentum generator takes
here values of one sign that coincides with that of the magnetic
field.

The two non-critical phases have essentially
different properties. Angular momentum takes half-integer values of
both signs in the sub-critical phase, but it takes half-integer values
of one sign only (correlated to the sign of the magnetic field) in the
super-critical phase. In both of these phases canonical \emph{vector}
coordinates and momenta can be constructed from the initial
non-commuting coordinates and momenta. In the sub-critical phase such
operators are hermitian, but in the super-critical phase they are not
hermitian. In the sub-critical phase the bosonic part of the super-conformal
symmetry is described by the non-compact $so(2,1)\sim su(1,1)$
algebra. In the super-critical phase it is changed into the compact
$so(3)\sim su(2)$ algebra.

When we try to unify the two-fold central extension of the
superextended Newton-Hooke symmetry with super-conformal symmetry,
the structure coefficients of the symmetry superalgebra transform
into certain functions depending on central charges. Linear,
Lie-superalgebraic structure can be achieved via the change of the
basis with coefficients depending on the noncommutativity
parameter $\theta$. The resulting Lie superalgebraic structure has
only one central charge.

In Ref. \cite{Alvarez:2007fw}, it was shown that the exotic
Newton-Hooke symmetry with associated coordinate non-commutativity
can be obtained from relativistic AdS${}_3$ via a certain
Wigner-In\"on\"u contraction. It would be interesting to extend the
analysis of  conformal and super- symmetries to the
context of AdS/CFT correspondence \cite{Balasubramanian:2008dm}, and  to
noncommutative fields \cite{CCGM}.

\vskip 0.2cm\noindent {\bf Acknowledgements}. P.A. and M.P. thank
the Department of Theoretical Physics of Zaragoza University
(Spain), and P.H. is indebted to the  Departamento de F\'\i sica
of Santiago University (Chile) for hospitality.
 The work has
been supported in part by CONICYT, FONDECYT Project 1050001, and
MeceSup Project FSM0605 (Chile), and by CICYT (grant
FPA2006-02315) and DGIID-DGA (grant 2008-E24/2) (Spain).

\end{document}